\documentclass[aps,prd,twocolumn, superscriptaddress, showkeys]{revtex4-1}

\usepackage{multirow}
\usepackage{graphicx}  
\usepackage{dcolumn}   
\usepackage{bm}        
\usepackage{amssymb}   
\usepackage{graphicx, epsfig, subfigure}
\usepackage{hyperref}
\usepackage[mathlines]{lineno}
\usepackage{color}
\usepackage{float}
\usepackage{enumerate}
\usepackage{enumitem}
\usepackage{tabularx}
\usepackage{multirow}
\usepackage{amsmath}
\usepackage{tensor}
\usepackage[export]{adjustbox}
\usepackage{upgreek} 

\usepackage{numcompress}
\newcommand{\gev}{GeV/$c$}

\newcommand{\mevtwo}{MeV/$c^{2}$}

\newcommand{\jpsi}{\ensuremath{J/\psi}}
\newcommand{\pp}{$p$+$p$}

\newcommand{\pT}{\ensuremath{p_\mathrm{T}}}

\newcommand{\sNN}{$\sqrt{s_{_\mathrm{NN}}}$}
\newcommand{\auau}{Au+Au}
\newcommand{\raa}{$R_{\mathrm{AA}}$}
\newcommand{\ups}{$\Upsilon$}
\newcommand{\npart}{$N_{\rm{part}}$}
\newcommand{\ncoll}{$N_{\rm{coll}}$}

\begin{document}

    \title{Measurement of sequential \ups\ suppression in Au+Au collisions at \sNN\ = 200 GeV with the STAR experiment} 

\affiliation{Abilene Christian University, Abilene, Texas   79699}
\affiliation{AGH University of Science and Technology, FPACS, Cracow 30-059, Poland}
\affiliation{Argonne National Laboratory, Argonne, Illinois 60439}
\affiliation{American University of Cairo, New Cairo 11835, New Cairo, Egypt}
\affiliation{Ball State University, Muncie, Indiana, 47306}
\affiliation{Brookhaven National Laboratory, Upton, New York 11973}
\affiliation{University of Calabria \& INFN-Cosenza, Italy}
\affiliation{University of California, Berkeley, California 94720}
\affiliation{University of California, Davis, California 95616}
\affiliation{University of California, Los Angeles, California 90095}
\affiliation{University of California, Riverside, California 92521}
\affiliation{Central China Normal University, Wuhan, Hubei 430079 }
\affiliation{University of Illinois at Chicago, Chicago, Illinois 60607}
\affiliation{Creighton University, Omaha, Nebraska 68178}
\affiliation{Czech Technical University in Prague, FNSPE, Prague 115 19, Czech Republic}
\affiliation{Technische Universit\"at Darmstadt, Darmstadt 64289, Germany}
\affiliation{ELTE E\"otv\"os Lor\'and University, Budapest, Hungary H-1117}
\affiliation{Frankfurt Institute for Advanced Studies FIAS, Frankfurt 60438, Germany}
\affiliation{Fudan University, Shanghai, 200433 }
\affiliation{University of Heidelberg, Heidelberg 69120, Germany }
\affiliation{University of Houston, Houston, Texas 77204}
\affiliation{Huzhou University, Huzhou, Zhejiang  313000}
\affiliation{Indian Institute of Science Education and Research (IISER), Berhampur 760010 , India}
\affiliation{Indian Institute of Science Education and Research (IISER) Tirupati, Tirupati 517507, India}
\affiliation{Indian Institute Technology, Patna, Bihar 801106, India}
\affiliation{Indiana University, Bloomington, Indiana 47408}
\affiliation{Institute of Modern Physics, Chinese Academy of Sciences, Lanzhou, Gansu 730000 }
\affiliation{University of Jammu, Jammu 180001, India}
\affiliation{Kent State University, Kent, Ohio 44242}
\affiliation{University of Kentucky, Lexington, Kentucky 40506-0055}
\affiliation{Lawrence Berkeley National Laboratory, Berkeley, California 94720}
\affiliation{Lehigh University, Bethlehem, Pennsylvania 18015}
\affiliation{Max-Planck-Institut f\"ur Physik, Munich 80805, Germany}
\affiliation{Michigan State University, East Lansing, Michigan 48824}
\affiliation{National Institute of Science Education and Research, HBNI, Jatni 752050, India}
\affiliation{National Cheng Kung University, Tainan 70101 }
\affiliation{Nuclear Physics Institute of the CAS, Rez 250 68, Czech Republic}
\affiliation{Ohio State University, Columbus, Ohio 43210}
\affiliation{Institute of Nuclear Physics PAN, Cracow 31-342, Poland}
\affiliation{Panjab University, Chandigarh 160014, India}
\affiliation{Purdue University, West Lafayette, Indiana 47907}
\affiliation{Rice University, Houston, Texas 77251}
\affiliation{Rutgers University, Piscataway, New Jersey 08854}
\affiliation{Universidade de S\~ao Paulo, S\~ao Paulo, Brazil 05314-970}
\affiliation{University of Science and Technology of China, Hefei, Anhui 230026}
\affiliation{South China Normal University, Guangzhou, Guangdong 510631}
\affiliation{Shandong University, Qingdao, Shandong 266237}
\affiliation{Shanghai Institute of Applied Physics, Chinese Academy of Sciences, Shanghai 201800}
\affiliation{Southern Connecticut State University, New Haven, Connecticut 06515}
\affiliation{State University of New York, Stony Brook, New York 11794}
\affiliation{Instituto de Alta Investigaci\'on, Universidad de Tarapac\'a, Arica 1000000, Chile}
\affiliation{Temple University, Philadelphia, Pennsylvania 19122}
\affiliation{Texas A\&M University, College Station, Texas 77843}
\affiliation{University of Texas, Austin, Texas 78712}
\affiliation{Tsinghua University, Beijing 100084}
\affiliation{University of Tsukuba, Tsukuba, Ibaraki 305-8571, Japan}
\affiliation{United States Naval Academy, Annapolis, Maryland 21402}
\affiliation{Valparaiso University, Valparaiso, Indiana 46383}
\affiliation{Variable Energy Cyclotron Centre, Kolkata 700064, India}
\affiliation{Warsaw University of Technology, Warsaw 00-661, Poland}
\affiliation{Wayne State University, Detroit, Michigan 48201}
\affiliation{Yale University, New Haven, Connecticut 06520}

\author{B.~E.~Aboona}\affiliation{Texas A\&M University, College Station, Texas 77843}
\author{J.~Adam}\affiliation{Czech Technical University in Prague, FNSPE, Prague 115 19, Czech Republic}
\author{L.~Adamczyk}\affiliation{AGH University of Science and Technology, FPACS, Cracow 30-059, Poland}
\author{J.~R.~Adams}\affiliation{Ohio State University, Columbus, Ohio 43210}
\author{I.~Aggarwal}\affiliation{Panjab University, Chandigarh 160014, India}
\author{M.~M.~Aggarwal}\affiliation{Panjab University, Chandigarh 160014, India}
\author{Z.~Ahammed}\affiliation{Variable Energy Cyclotron Centre, Kolkata 700064, India}
\author{D.~M.~Anderson}\affiliation{Texas A\&M University, College Station, Texas 77843}
\author{E.~C.~Aschenauer}\affiliation{Brookhaven National Laboratory, Upton, New York 11973}
\author{J.~Atchison}\affiliation{Abilene Christian University, Abilene, Texas   79699}
% ...

\collaboration{STAR Collaboration}\noaffiliation


\date{\today}

\begin{abstract}
We report on measurements of sequential \ups\ suppression in Au+Au collisions at \sNN\ = 200 GeV with the STAR detector at the Relativistic Heavy Ion Collider (RHIC) through both the dielectron and dimuon decay channels. In the 0-60\% centrality class, the nuclear modification factors (\raa), which quantify the level of yield suppression in heavy-ion collisions compared to \pp\ collisions, for \ups(1S) and \ups(2S) are $0.40 \pm 0.03~\textrm{(stat.)} \pm 0.03~\textrm{(sys.)} \pm 0.09~\textrm{(norm.)}$ and $0.26 \pm 0.08~\textrm{(stat.)} \pm 0.02~\textrm{(sys.)} \pm 0.06~\textrm{(norm.)}$, respectively, while the upper limit of the \ups(3S)\ \raa\ is 0.17 at a 95\% confidence level. This provides experimental evidence that the \ups(3S) is significantly more suppressed than the \ups(1S) at RHIC. The level of suppression for \ups(1S) is comparable to that observed at the much higher collision energy at the Large Hadron Collider. These results point to the creation of a medium at RHIC whose temperature is sufficiently high to strongly suppress excited \ups\ states.
\end{abstract}

\maketitle


A primary goal of the Relativistic Heavy Ion Collider (RHIC) is to create and study the properties of the Quark-Gluon Plasma (QGP)~\cite{STAR_WhitePaper, PHENIX_WhitePaper, PHOBOS_WhitePaper, BRAHMS_WhitePaper}. Quantum chromodynamics (QCD) predicts that the confining potential of a heavy quark-antiquark pair is color-screened in the QGP \cite{colorscreen}, leading to the dissociation of quarkonium states. Such a static dissociation is expected to happen when the quarkonium state size is larger than the Debye screening length of the medium \cite{Satz:1983jp}, which is inversely proportional to the medium temperature. In addition, dynamical dissociation, arising from inelastic scatterings between quarkonia and medium constituents, can also lead to quarkoninum breakup, whose impact becomes more profound with increasing medium temperature and for quarkonia of larger sizes~\cite{Laine:2006ns, Burnier:2015tda, Chen:2017jje}. Consequently, quarkonium states of different sizes suffer from different levels of suppression in the QGP (``sequential suppression") compared to the vacuum expectation \cite{Digal:2001ue, Mocsy:2007jz, Burnier:2015tda}. Heavy quarkonia are therefore considered promising probes to study the color deconfinement, in-medium QCD force, and the QGP's thermodynamic properties \cite{Rothkopf:2019ipj}. 

In heavy-ion collisions, sequential suppression of charmonium states has been observed, with the yield of the larger $\psi$(2S) mesons further reduced compared to \jpsi~\cite{NA50:2006yzz, ALICE:2015jrl, CMS:2016wgo, CMS:2017uuv, ATLAS:2018hqe, ALICE:2022jeh}. Compared to charmonia, bottomonia (\ups(1S), \ups(2S), and \ups(3S)), with \ups(1S) being the smallest in size and \ups(3S) the biggest, provide a longer lever arm in probing the QGP. According to lattice QCD calculations based on a complex quark-antiquark potential, the span of the dissociation temperature for the three bottomonium states is about a factor of four larger than that for the two charmonium states \cite{Burnier:2015tda}. Furthermore, bottomonia are considered cleaner probes than charmonia since the regeneration contribution, originating from deconfined heavy quark-antiquark pairs combining into quarkonium states, is expected to be smaller for bottomonia due to the smaller production cross section of $b\bar{b}$ quarks~\cite{Zhao:2010nk, Du:2017qkv}. When interpreting \ups\ measurements in heavy-ion collisions, Cold Nuclear Matter (CNM) effects, arising from the presence of nuclei in the collision but not related to the QGP, need to be considered \cite{Ferreiro:2009ur, Arleo:2014oha, Gavin:1996yd}. The CNM effects can be quantified through measurements of \ups\ production in $d$+Au collisions at RHIC \cite{STAR:2013kwk}, which show a hint of suppression for the three \ups\ states combined. 

Sequential suppression of the three \ups\ states has been observed in Pb+Pb collisions at the LHC~\cite{CMS:2016rpc, CMS:2018zza, ALICE:2020wwx}. In Au+Au collisions at the center-of-mass energy per nucleon-nucleon pair (\sNN) of 200 GeV \cite{STAR:2013kwk} and U+U collisions at \sNN\ = 193 GeV~\cite{Adamczyk:2016dzv} at RHIC, previous measurements revealed a hint of stronger suppression for \ups(2S+3S) compared to \ups(1S) with a significance of less than 1.5$\sigma$. To fully utilize the constraining power of quarkonium sequential suppression on the QGP's temperature profile and modifications to the QCD force in the QGP \cite{Rothkopf:2019ipj} at RHIC, differential measurements of ground and excited \ups\ states separately with improved precision are crucially needed. 

In this Letter, we report the latest measurements of the suppression of \ups(1S), \ups(2S) and \ups(3S) production in Au+Au collisions at \sNN\ = 200 GeV. \ups\ mesons are reconstructed through both dielectron and dimuon decay decay channels. The suppression is quantified with the nuclear modification factor ($R_{\textrm{AA}}$), which is the ratio of the quarkonium yield measured in nucleus-nucleus (A+A) collisions to that in \pp\ collisions, scaled by the average number of binary nucleon-nucleon collisions (\ncoll). Results are presented as a function of the collision centrality and the \ups\ transverse momentum (\pT), where central (peripheral) collisions correspond to incoming nuclei most (least) overlapping with each other.

Subsystems of the STAR experiment~\cite{Ackermann:2002ad} relevant for this analysis are the Time Projection Chamber (TPC)~\cite{Anderson:2003ur}, the Barrel Electromagnetic Calorimeter (BEMC)~\cite{Beddo:2002zx} and the Muon Telescope Detector (MTD)~\cite{Ruan:2009ug, STAR:2019fge}. The TPC is used for track reconstruction and particle identification (PID), while the BEMC and MTD are used for triggering on and identifying electrons and muons, respectively. The TPC and the BEMC have a full azimuthal coverage within the pseudorapidity range of $|\eta| < 1$. The MTD covers about 45\% in azimuth within $|\eta| < 0.5$. The $\Upsilon \rightarrow e^+e^-$ analysis is performed on a data set of Au+Au collisions corresponding to an integrated luminosity of 2.3 nb$^{-1}$, which was collected in 2011 with the BEMC trigger requiring the presence of a single tower with transverse energy deposition above 3.5 GeV. Electrons with $\pT > 3.5$ GeV/$c$ are selected based on their ionization energy loss ($dE/dx$) measured in the TPC. Cuts on the ratio of energy deposition in BEMC over associated track momentum ($E/p$), and on the position differences along beam and azimuthal directions between matched BEMC tower and TPC track are applied to further reject hadrons. For the $\Upsilon \rightarrow \mu^+\mu^-$ analysis, a sample of Au+Au collisions, recorded with the MTD dimuon trigger in 2014 and 2016 and corresponding to an integrated luminosity of 27 nb$^{-1}$, is utilized. The dimuon trigger requires the presence of two muon candidates, identified based on the particles' flight time, in the MTD. The leading muon is required to have \pT\ above 4 GeV/$c$ and the sub-leading above 1.5 GeV/$c$. Besides $dE/dx$, muon candidates are identified utilizing position and timing information measured by the MTD \cite{Huang:2016dbm, STAR:2019fge}.

A Glauber model simulation is used for centrality classification \cite{Miller:2007ri}. The charged-particle multiplicity distribution within $|\eta|<0.5$ obtained from the simulation is matched to the measured one at large multiplicity values. The average number of participating nucleons (\npart) and \ncoll\ are calculated for each centrality class, and their uncertainties are evaluated by varying different components of the Glauber model. Data are divided into three centrality bins: 0-10\%, 10-30\%, and 30-60\%, as well as three \ups\ \pT\ bins: 0-2 \gev, 2-5 \gev, and 5-10 \gev.

\begin{figure}[htbp]
\includegraphics[width=0.48\textwidth]{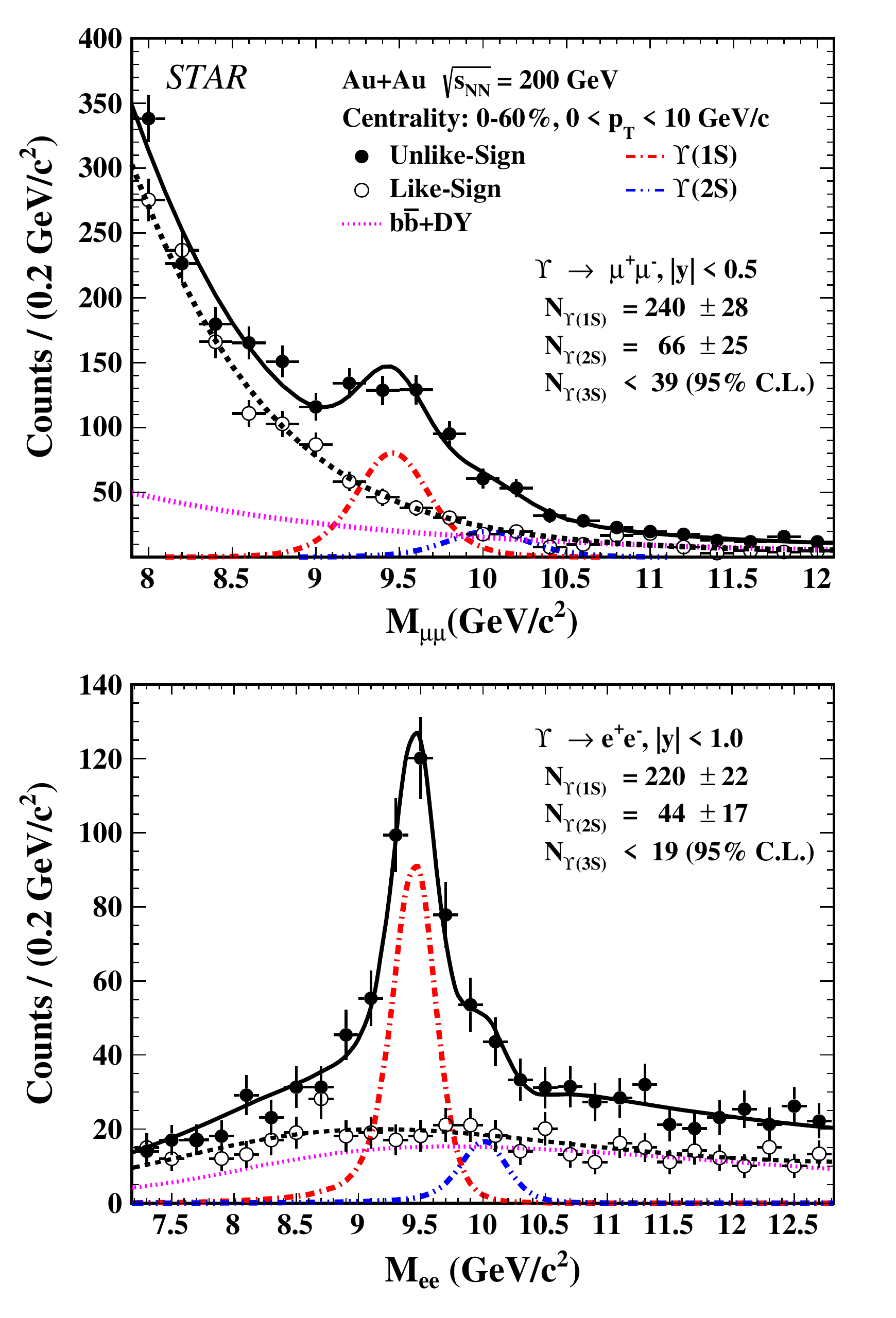} 
\caption{Invariant mass distributions of \ups\ candidates for $0 < \pT < 10$ GeV/$c$ reconstructed via the dimuon decay channel within $|y|<0.5$ (top) and the dielectron decay channel within $|y| < 1$ (bottom). Unlike-sign and like-sign distributions are shown as full and open circles, respectively. Solid lines are fits to the unlike-sign distributions, while lines of other styles represent individual components included in the fit. See more details in the text.}\label{fig:fig1}
\end{figure}
 
The invariant mass spectra of the \ups\ candidates are reconstructed via the dimuon decay channel within the rapidity range of $|y| < 0.5$ and via the dielectron decay channel within $|y| < 1$. Figure~\ref{fig:fig1} shows the unlike-sign lepton-pair distributions (full circles), along with like-sign ones (open circles) which are used for determining the shape and magnitude of the combinatorial background. An unbinned maximum-likelihood fit is performed simultaneously on the unlike-sign and like-sign distributions to obtain the raw yields for the three \ups\ states. The lineshapes of the \ups\ mass peaks are determined from GEANT3 simulations \cite{Brun:1987ma} of the STAR detector, in which the $\Upsilon \rightarrow \mu^+\mu^-$ or $\Upsilon \rightarrow e^+e^-$ decays are embedded into Au+Au collision events, and reconstructed in the same way as real data. The track momentum resolution in the simulation is further tuned to match the $J/\psi$ width as a function of \pT\ reconstructed using the same Au+Au data. The \ups(1S) peak widths are 221 \mevtwo\ and 129 \mevtwo\ for the dimuon and dielectron decay channels, respectively. The shape of the correlated background from $b\bar{b}$ decays and Drell-Yan processes is determined with PYTHIA6 simulations \cite{Sjostrand:2006za} incorporating realistic detector response, while its yield is left as a free fit parameter. With current statistics, no \ups(3S) signal is observed in either decay channel, and therefore only the upper limits of \ups(3S) yields are estimated with the Feldman-Cousins method \cite{Feldman:1997qc} at a 95\% confidence level.

The TPC acceptance and tracking efficiency are determined based on aforementioned embedding sample. In the $\Upsilon \rightarrow e^+e^-$ analysis, the BEMC trigger efficiency is evaluated using the same embedding sample while the electron PID efficiency is estimated using a pure electron sample from photon conversions in real data. In the $\Upsilon \rightarrow \mu^+\mu^-$ analysis, a pure muon sample from $J/\psi$ decays is used to evaluate the muon PID efficiencies based on $dE/dx$ and the MTD timing information. The embedding sample is used to estimate the additional PID efficiency related to using the MTD position information, and the MTD  acceptance. The MTD response efficiency, referring to the probability for a muon to generate a signal in the MTD when hitting its active volume, is obtained from cosmic-ray data \cite{STAR:2019fge}. The MTD trigger efficiency, {\it i.e.} the fraction of muons surviving the trigger cut on the flight time, is evaluated based on the flight time distribution extracted from the \pp\ data taken in 2015. Since the MTD occupancy is very low even in 0-10\% central Au+Au collisions, the multiplicity difference between \pp\ and Au+Au collisions is irrelevant for this purpose \cite{STAR:2019fge}.

Several sources of systematic uncertainty are considered. Variations in the signal extraction procedure, including fit range, lineshapes of the mass peaks, combinatorial and residual background shapes, are made and the Root Mean Square (RMS) of these variations is taken as the systematic uncertainty. For the dielectron (dimuon) analysis, the resulting uncertainty ranges between 1.7-4.2\% (1.5-4.0\%) and 2.1-8.3\% (1.7-98\%) for \ups(1S) and \ups(2S) in different centrality and \pT\ bins, and is 2.3 (4.9) in absolute value for \ups(3S) yield integrated over \pT\ in 0-60\% centrality. Another major source of uncertainty arises from efficiency corrections. For efficiencies evaluated based on the embedding sample, their uncertainties are estimated by varying cuts in data analysis and simulation simultaneously, correcting the raw yields, and taking the RMS of the variations in the corrected yield as the uncertainty. For efficiencies evaluated using data-driven methods, statistical errors of the data samples are treated as systematic uncertainties. Uncertainties in MTD response and trigger efficiencies are estimated using the same method as in \cite{STAR:2019fge}. The overall efficiency uncertainties apply equally to all three \ups\ states, and they vary from 3.7\% to 19.8\% (11.6\% to 18.6\%) depending on centrality and \pT\ for the dielectron (dimuon) analysis. Finally, the individual sources are added in quadrature to obtain the total systematic uncertainties for the \ups\ yields. When combining the dimuon and dielectron results, the TPC tracking efficiency uncertainties are treated as fully correlated while all other uncertainties are uncorrelated. 

The reference \ups(1S+2S+3S) production cross section in \pp\ collisions at the center-of-mass energy ($\sqrt{s}$) of 200 GeV is \mbox{$\frac{d\sigma}{dy}|_{|y|<0.5} = 75 \pm 15$ pb}, obtained by combining STAR and PHENIX measurements \cite{Abelev:2010am, STAR:2013kwk, PHENIX:2014tbe}. The cross sections of individual \ups\ states are calculated based on the total cross section and their yield ratios from world data \cite{Zha:2013uoa}. To obtain the reference cross sections in different \pT\ bins, the measured \ups\ \pT\ spectra at different collision energies \cite{NuSea:2007ult, CDF:2001fdy, CMS:2016rpc, CMS:2013qur} are parameterized with the functional form $C\times\pT/(e^{\pT/T}+1)$ \cite{Adamczyk:2016dzv}, where $C$ is a normalization factor and $T$ is the shape parameter. The dependence of $T$ on $\log(\sqrt{s})$ is fit with both a linear and a power-law function, and the average interpolated $T$ values at $\sqrt{s}$ = 200 GeV from the two fits, $i.e.$, $1.40\pm0.06$ \gev\ and $1.51\pm0.10$ \gev\ for \ups(1S) and \ups(2S), are obtained. Systematic uncertainties arise from the uncertainties on the measured \ups\ spectra and the functional form used for interpolation.

The \raa\ of individual \ups\ states in \auau\ collisions at \sNN\ = 200 GeV is obtained by combining results from dimuon and dielectron decay channels using the inverse of the quadratic sum of statistical errors and uncorrelated systematic uncertainties as weights, since the results from the two analyses are consistent despite the different rapidity coverages. Similarly, no strong dependence of \ups\ \raa\ on rapidity within $|y|<1$ is observed at the LHC \cite{CMS:2018zza}. 

\begin{figure}[t]
\includegraphics[width=0.48\textwidth]{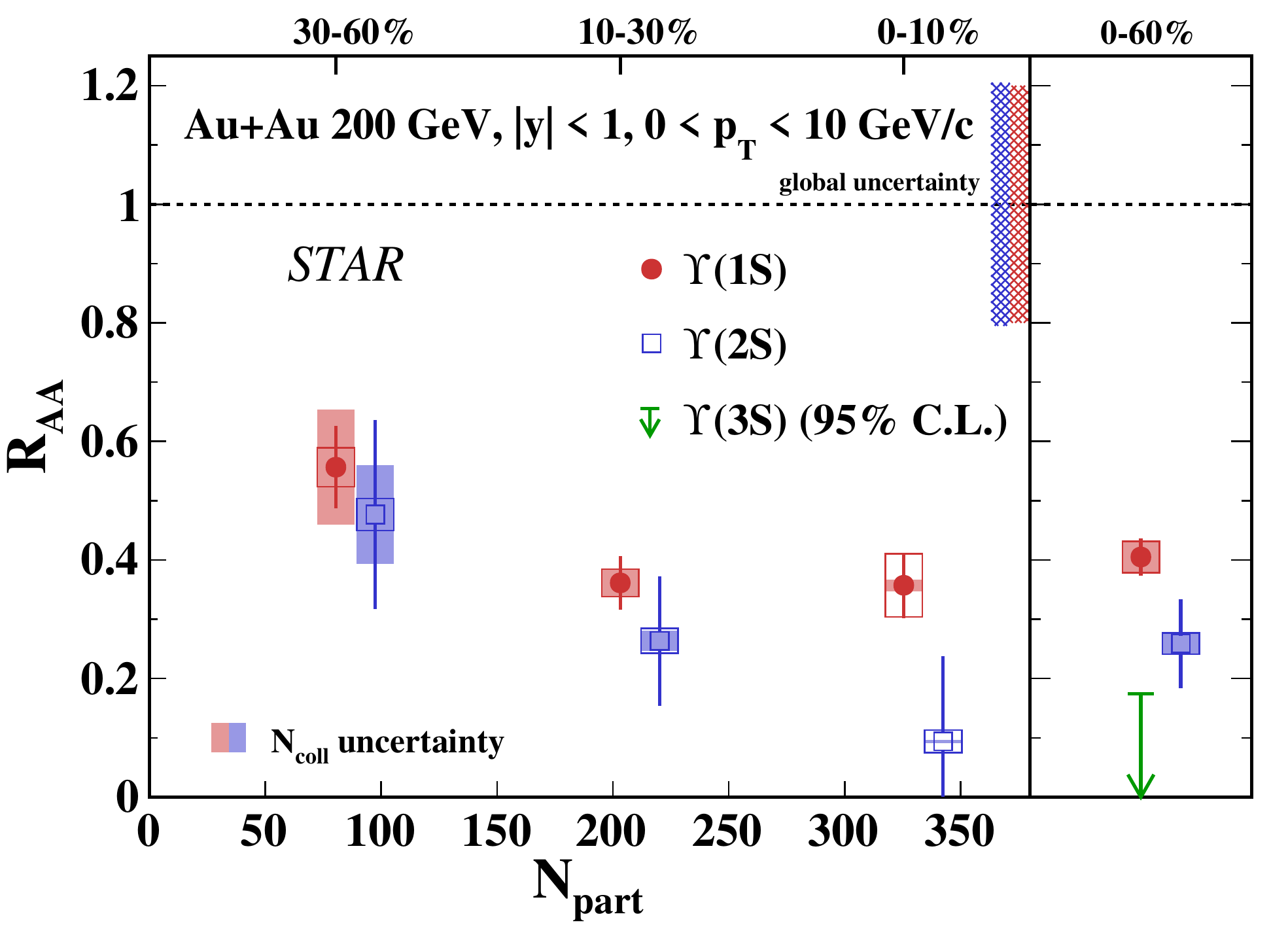} 
\caption{Left: \ups(1S) (circles) and \ups(2S) (squares) \raa\ as a function of $N_{\textrm{part}}$ for $\pT < 10$ \gev. Data points for \ups(2S) are displaced horizontally for better visibility. The vertical bars on data points indicate statistical errors, while the systematic uncertainties are shown as boxes. Shadowed bands around each marker depict the systematic uncertainties from \ncoll. The bands at unity indicate the global uncertainties. Right: \raa\ for various \ups\ states, including the 95\% upper limit for \ups(3S), in 0-60\% Au+Au collisions.}
\label{fig:fig2}
\end{figure}

Figure~\ref{fig:fig2} shows the \raa\ of \ups(1S) and \ups(2S) as a function of \npart\ in three centrality intervals. The global uncertainties, shown as bands at unity and fully correlated among different \ups\ states, originate from the relative uncertainties of the reference \pp\ yields.
Both \ups(1S) and \ups(2S) are suppressed in all three centrality intervals with a hint of increasing suppression from the 30-60\% to the 0-10\% centrality bin, consistent with the expected increasing hot medium effect towards central collisions. In the 0-60\% centrality class, the upper limit of the \ups(3S) \raa\ with a 95\% confidence level is estimated to be 0.17. \ups(3S) is significantly more suppressed than \ups(1S), given that even the upper limit of \ups(3S) \raa\ at a 99\% confidence level, \textit{i.e.}\ 0.26, is still lower than the \ups(1S) \raa\ of $0.40 \pm 0.03~\textrm{(stat.)} \pm 0.03~\textrm{(sys.)} \pm 0.09~\textrm{(norm.)}$. Here, the normalization uncertainty includes uncertainties in \pp\ reference and \ncoll. A hint is seen that the level of suppression for \ups(2S), whose \raa\ is $0.26 \pm 0.08~\textrm{(stat.)} \pm 0.02~\textrm{(sys.)} \pm 0.06~\textrm{(norm.)}$, is between \ups(1S) and \ups(3S). These results are consistent with a sequential suppression pattern, similar to that observed at the LHC~\cite{CMS:2018zza}. 

\begin{figure}[tp]
\includegraphics[width=0.48\textwidth]{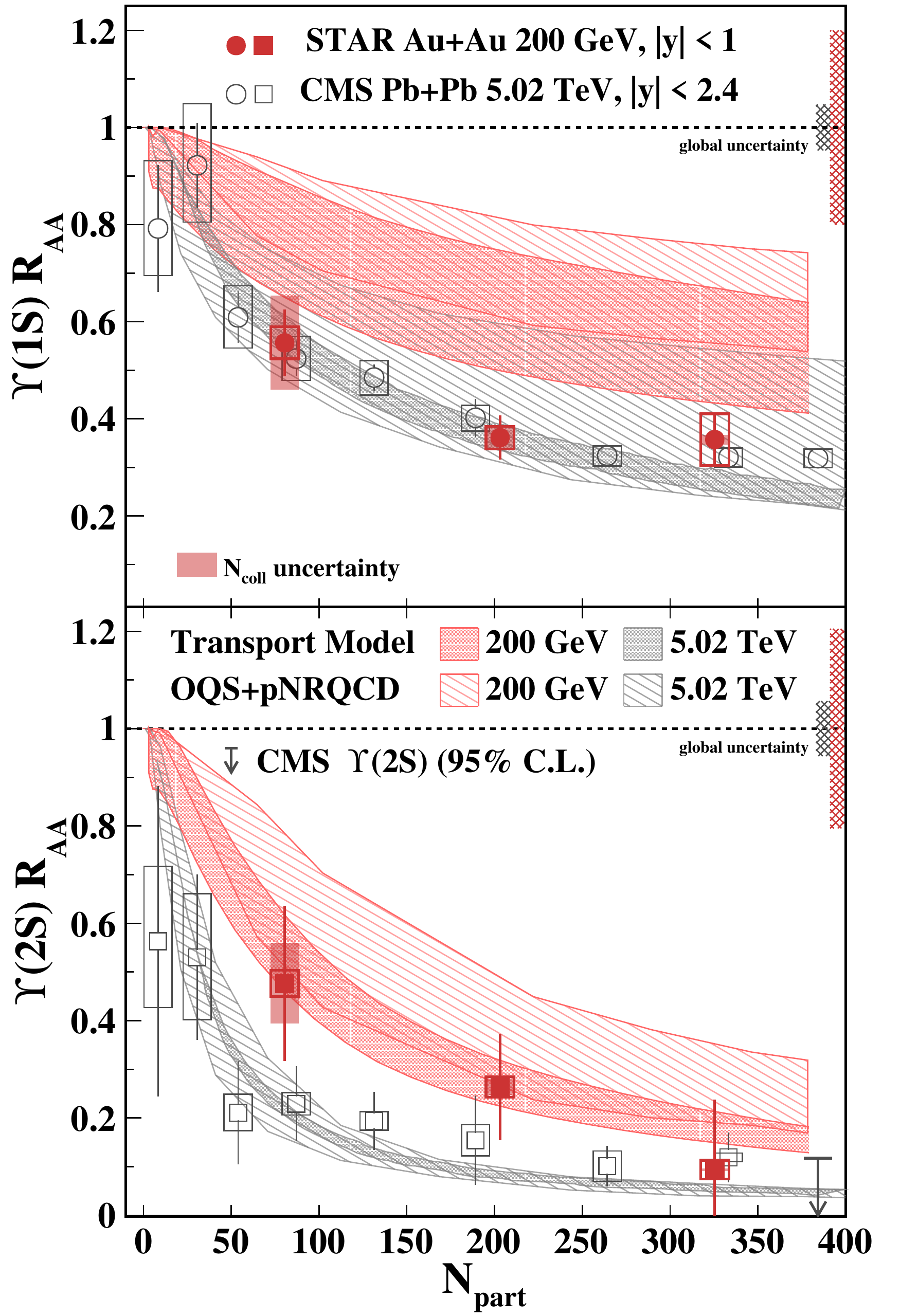} 
\caption{\ups(1S) (top) and \ups(2S) (bottom) \raa\ as a function of $N_{\textrm{part}}$ for $\pT < 10$ \gev, compared to similar measurements in Pb+Pb collisions at \sNN\ = 5.02 TeV (open symbols), as well as model calculations (bands). The two bands at unity indicate the global uncertainties with the left one for CMS and the right one for STAR.}\label{fig:fig3}
\end{figure}

The \auau\ results are compared to similar measurements in Pb+Pb collisions at \sNN\ = 5.02 TeV~\cite{CMS:2018zza} in Fig.~\ref{fig:fig3}.
\ups(1S) exhibits a similar magnitude of suppression at the two collision energies that differ by about a factor of 25, while there is a hint that the \ups(2S) might be less suppressed at RHIC in peripheral collisions even though the STAR and CMS measurements are consistent within uncertainties. It is plausible that the suppression of inclusive \ups(1S) arises mainly from the suppression of excited states that feed down to \ups(1S) \cite{Lansberg:2019adr} and the CNM effects \cite{STAR:2013kwk,ATLAS:2017prf,CMS:2022wfi}, while the primordial \ups(1S) are not significantly suppressed in the QGP in both 200 GeV Au+Au and 5.02 TeV Pb+Pb collisions. Figure~\ref{fig:fig3} also shows the comparison between data and two calculations based on Open Quantum System (OQS) plus potential Non-Relativistic QCD (pNRQCD)~\cite{Brambilla:2021wkt, Brambilla:2020qwo, Brambilla:2022ynh} and a transport model~\cite{Du:2017qkv}. The OQS+pNRQCD model solves a Lindblad equation for the evolution of the quarkonium reduced density matrix using the pNRQCD effective field theory \cite{Brambilla:2022ynh}. Correlated regeneration and feed-down contributions from excited states are included, but the CNM effects are not. Systematic uncertainties stem from variations in the transport coefficients suggested by lattice QCD calculations. The transport model employs a temperature-dependent binding energy, and uses a kinetic rate equation to simulate the time evolution of bottomonium abundances including dissociation and regeneration contributions. Both feed-down and CNM effects are taken into account, and the model uncertainties arise from the range of CNM effects guided by data \cite{STAR:2013kwk}. For the \ups(1S) $R_{\textrm{AA}}$, both models are consistent with the STAR and CMS measurements within uncertainties even though the STAR data seem to be systematically below the model calculations. For \ups(2S), model calculations are also consistent with data.

\begin{figure}[tp] 
\includegraphics[width=0.48\textwidth]{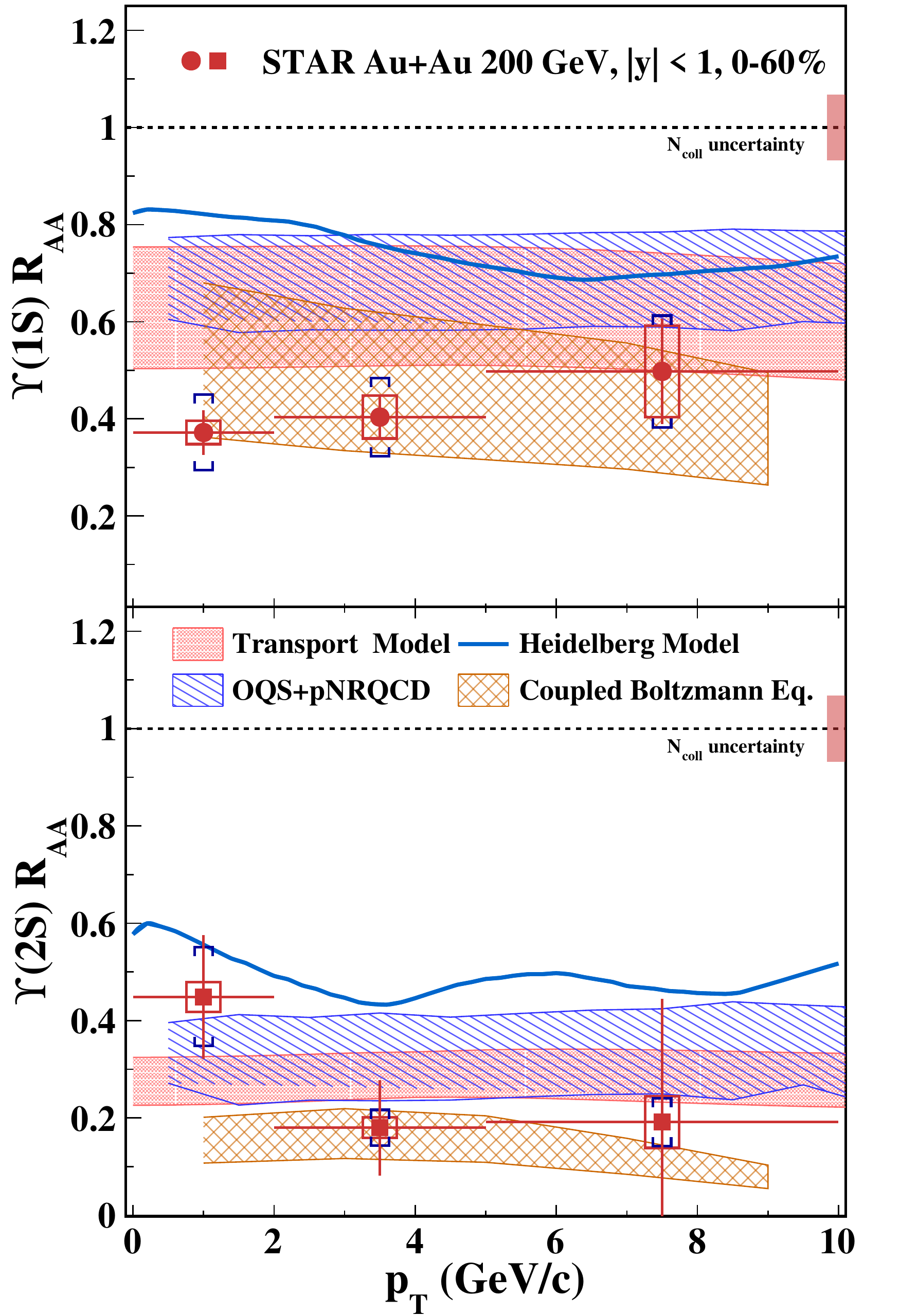} \centering
\caption{\ups(1S) (top) and \ups(2S) (bottom) \raa\ as a function of \pT\ in the 0-60\% centrality class, compared to different model calculations. The boxes and brackets around the data points represent systematic uncertainties from Au+Au analysis and \pp\ reference, respectively. The band at unity shows the uncertainty in \ncoll.}\label{fig:fig4}
\end{figure}

Figure~\ref{fig:fig4} shows the \raa\ for \ups(1S) and \ups(2S) as a function of \pT. No significant dependence on \pT\ is observed. The OQS+pNRQCD and transport model calculations, which predict little \pT\ dependence, are shown for comparison. The
measurements are also compared to a model that uses a set of coupled Boltzmann equations to simultaneously describe the in-medium evolution of heavy quarks and quarkonia in the QGP~\cite{Yao:2020xzw}. It incorporates elastic and inelastic scatterings of heavy quarks with medium constituents, as well as quarkonium dissociation and regeneration. The dominant uncertainty arises from the estimation of CNM effects. The model calculations are consistent with data within uncertainties. The Heidelberg model \cite{Hoelck:2016tqf}, which includes a QCD-inspired complex potential, an explicit treatment of gluon-induced dissociation and reduced feed-down from higher states, overshoots data, partly due to the lack of CNM effects.

In summary, we report the measurements of \ups\ production in Au+Au collisions at \sNN\ = 200 GeV via both the dielectron and dimuon decay channels with the STAR experiment. The \raa\ for \ups(1S) and \ups(2S) is measured as a function of collision centrality and \pT, while an upper limit is derived for the \ups(3S) \raa\ integrated over centrality and \pT. The results in the 0-60\% centrality class are consistent with the sequential suppression pattern, namely that the \ups(3S) is significantly more suppressed than the \ups(1S) and the \ups(2S) \raa\ lies between those of \ups(1S) and \ups(3S). No clear \pT\ dependence of the suppression is observed for \ups(1S) and \ups(2S). The magnitude of the \ups(1S) suppression at RHIC is comparable to that measured at the LHC. Model calculations are consistent with data within the uncertainties, although a larger \ups\ suppression is predicted at the LHC. Results presented in this paper can help further constrain model calculations on bottomonium suppression in heavy-ion collisions, and improve our understanding of the in-medium heavy quark-antiquark potential and thermodynamic properties of the QGP at RHIC.\\

We thank the RHIC Operations Group and RCF at BNL, the NERSC Center at LBNL, and the Open Science Grid consortium for providing resources and support.  This work was supported in part by the Office of Nuclear Physics within the U.S. DOE Office of Science, the U.S. National Science Foundation, National Natural Science Foundation of China, Chinese Academy of Science, the Ministry of Science and Technology of China and the Chinese Ministry of Education, the Higher Education Sprout Project by Ministry of Education at NCKU, the National Research Foundation of Korea, Czech Science Foundation and Ministry of Education, Youth and Sports of the Czech Republic, Hungarian National Research, Development and Innovation Office, New National Excellency Programme of the Hungarian Ministry of Human Capacities, Department of Atomic Energy and Department of Science and Technology of the Government of India, the National Science Centre and WUT ID-UB of Poland, the Ministry of Science, Education and Sports of the Republic of Croatia, German Bundesministerium f\"ur Bildung, Wissenschaft, Forschung and Technologie (BMBF), Helmholtz Association, Ministry of Education, Culture, Sports, Science, and Technology (MEXT) and Japan Society for the Promotion of Science (JSPS).


\bibliography{Paper_Ups_AuAuNotes}

\begin{thebibliography}{52}
\expandafter\ifx\csname natexlab\endcsname\relax\def\natexlab#1{#1}\fi
\providecommand{\url}[1]{\texttt{#1}}
\providecommand{\href}[2]{#2}
\providecommand{\path}[1]{#1}
\providecommand{\DOIprefix}{doi:}
\providecommand{\ArXivprefix}{arXiv:}
\providecommand{\URLprefix}{URL: }
\providecommand{\Pubmedprefix}{pmid:}
\providecommand{\doi}[1]{\href{http://dx.doi.org/#1}{\path{#1}}}
\providecommand{\Pubmed}[1]{\href{pmid:#1}{\path{#1}}}
\providecommand{\bibinfo}[2]{#2}
\ifx\xfnm\relax \def\xfnm[#1]{\unskip,\space#1}\fi
\bibitem[{Adams et~al.(2005)}]{STAR_WhitePaper}
\bibinfo{author}{J.~Adams}, et~al. (\bibinfo{collaboration}{STAR}),
\newblock \bibinfo{title}{{Experimental and theoretical challenges in the
  search for the quark gluon plasma: The STAR Collaboration's critical
  assessment of the evidence from RHIC collisions}},
\newblock \bibinfo{journal}{Nucl. Phys. A} \bibinfo{volume}{757}
  (\bibinfo{year}{2005}) \bibinfo{pages}{102--183}.
\bibitem[{Adcox et~al.(2005)}]{PHENIX_WhitePaper}
\bibinfo{author}{K.~Adcox}, et~al. (\bibinfo{collaboration}{PHENIX}),
\newblock \bibinfo{title}{{Formation of dense partonic matter in relativistic
  nucleus-nucleus collisions at RHIC: Experimental evaluation by the PHENIX
  collaboration}},
\newblock \bibinfo{journal}{Nucl. Phys. A} \bibinfo{volume}{757}
  (\bibinfo{year}{2005}) \bibinfo{pages}{184--283}.
\bibitem[{Back et~al.(2005)}]{PHOBOS_WhitePaper}
\bibinfo{author}{B.~Back}, et~al. (\bibinfo{collaboration}{PHOBOS}),
\newblock \bibinfo{title}{{The PHOBOS perspective on discoveries at RHIC}},
\newblock \bibinfo{journal}{Nucl. Phys. A} \bibinfo{volume}{757}
  (\bibinfo{year}{2005}) \bibinfo{pages}{28--101}.
\bibitem[{Arsene et~al.(2005)}]{BRAHMS_WhitePaper}
\bibinfo{author}{I.~Arsene}, et~al. (\bibinfo{collaboration}{BRAHMS}),
\newblock \bibinfo{title}{{Quark gluon plasma and color glass condensate at
  RHIC? The Perspective from the BRAHMS experiment}},
\newblock \bibinfo{journal}{Nucl. Phys. A} \bibinfo{volume}{757}
  (\bibinfo{year}{2005}) \bibinfo{pages}{1--27}.
\bibitem[{Matsui and Satz(1986)}]{colorscreen}
\bibinfo{author}{T.~Matsui}, \bibinfo{author}{H.~Satz},
\newblock \bibinfo{title}{{$J/\psi$ Suppression by Quark-Gluon Plasma
  Formation}},
\newblock \bibinfo{journal}{Phys. Lett.} \bibinfo{volume}{B178}
  (\bibinfo{year}{1986}) \bibinfo{pages}{416}.
\bibitem[{Satz(1984)}]{Satz:1983jp}
\bibinfo{author}{H.~Satz},
\newblock \bibinfo{title}{{Color Screening in SU($N$) Gauge Theory at Finite
  Temperature}},
\newblock \bibinfo{journal}{Nucl. Phys. A} \bibinfo{volume}{418}
  (\bibinfo{year}{1984}) \bibinfo{pages}{447C--465C}.
\bibitem[{Laine et~al.(2007)Laine, Philipsen, Romatschke, and
  Tassler}]{Laine:2006ns}
\bibinfo{author}{M.~Laine}, \bibinfo{author}{O.~Philipsen},
  \bibinfo{author}{P.~Romatschke}, \bibinfo{author}{M.~Tassler},
\newblock \bibinfo{title}{{Real-time static potential in hot QCD}},
\newblock \bibinfo{journal}{JHEP} \bibinfo{volume}{03} (\bibinfo{year}{2007})
  \bibinfo{pages}{054}.
\bibitem[{Burnier et~al.(2015)Burnier, Kaczmarek, and
  Rothkopf}]{Burnier:2015tda}
\bibinfo{author}{Y.~Burnier}, \bibinfo{author}{O.~Kaczmarek},
  \bibinfo{author}{A.~Rothkopf},
\newblock \bibinfo{title}{{Quarkonium at finite temperature: Towards realistic
  phenomenology from first principles}},
\newblock \bibinfo{journal}{JHEP} \bibinfo{volume}{12} (\bibinfo{year}{2015})
  \bibinfo{pages}{101}.
\bibitem[{Chen and He(2017)}]{Chen:2017jje}
\bibinfo{author}{S.~Chen}, \bibinfo{author}{M.~He},
\newblock \bibinfo{title}{{Gluo-dissociation of heavy quarkonium in the
  quark-gluon plasma reexamined}},
\newblock \bibinfo{journal}{Phys. Rev. C} \bibinfo{volume}{96}
  (\bibinfo{year}{2017}) \bibinfo{pages}{034901}.
\bibitem[{Digal et~al.(2001)Digal, Petreczky, and Satz}]{Digal:2001ue}
\bibinfo{author}{S.~Digal}, \bibinfo{author}{P.~Petreczky},
  \bibinfo{author}{H.~Satz},
\newblock \bibinfo{title}{{Quarkonium feed down and sequential suppression}},
\newblock \bibinfo{journal}{Phys. Rev. D} \bibinfo{volume}{64}
  (\bibinfo{year}{2001}) \bibinfo{pages}{094015}.
\bibitem[{Mocsy and Petreczky(2007)}]{Mocsy:2007jz}
\bibinfo{author}{A.~Mocsy}, \bibinfo{author}{P.~Petreczky},
\newblock \bibinfo{title}{{Color screening melts quarkonium}},
\newblock \bibinfo{journal}{Phys. Rev. Lett.} \bibinfo{volume}{99}
  (\bibinfo{year}{2007}) \bibinfo{pages}{211602}.
\bibitem[{Rothkopf(2020)}]{Rothkopf:2019ipj}
\bibinfo{author}{A.~Rothkopf},
\newblock \bibinfo{title}{{Heavy Quarkonium in Extreme Conditions}},
\newblock \bibinfo{journal}{Phys. Rept.} \bibinfo{volume}{858}
  (\bibinfo{year}{2020}) \bibinfo{pages}{1--117}.
\bibitem[{Alessandro et~al.(2007)}]{NA50:2006yzz}
\bibinfo{author}{B.~Alessandro}, et~al. (\bibinfo{collaboration}{NA50}),
\newblock \bibinfo{title}{{psi-prime production in Pb-Pb collisions at
  158-GeV/nucleon}},
\newblock \bibinfo{journal}{Eur. Phys. J. C} \bibinfo{volume}{49}
  (\bibinfo{year}{2007}) \bibinfo{pages}{559--567}.
\bibitem[{Adam et~al.(2016)}]{ALICE:2015jrl}
\bibinfo{author}{J.~Adam}, et~al. (\bibinfo{collaboration}{ALICE}),
\newblock \bibinfo{title}{{Differential studies of inclusive
  J/\ensuremath{\psi} and \ensuremath{\psi}(2S) production at forward rapidity
  in Pb-Pb collisions at $ \sqrt{s_{\mathrm{NN}}}=2.76 $ TeV}},
\newblock \bibinfo{journal}{JHEP} \bibinfo{volume}{05} (\bibinfo{year}{2016})
  \bibinfo{pages}{179}.
\bibitem[{Sirunyan et~al.(2017)}]{CMS:2016wgo}
\bibinfo{author}{A.~M. Sirunyan}, et~al. (\bibinfo{collaboration}{CMS}),
\newblock \bibinfo{title}{{Relative Modification of Prompt
  \ensuremath{\psi}(2S) and J/\ensuremath{\psi} Yields from pp to PbPb
  Collisions at $\sqrt{s_{_\mathrm{NN}}}$ = 5.02 TeV}},
\newblock \bibinfo{journal}{Phys. Rev. Lett.} \bibinfo{volume}{118}
  (\bibinfo{year}{2017}) \bibinfo{pages}{162301}.
\bibitem[{Sirunyan et~al.(2018)}]{CMS:2017uuv}
\bibinfo{author}{A.~M. Sirunyan}, et~al. (\bibinfo{collaboration}{CMS}),
\newblock \bibinfo{title}{{Measurement of prompt and nonprompt charmonium
  suppression in $\text {PbPb}$ collisions at 5.02 $\,\text {Te}\text {V}$}},
\newblock \bibinfo{journal}{Eur. Phys. J. C} \bibinfo{volume}{78}
  (\bibinfo{year}{2018}) \bibinfo{pages}{509}.
\bibitem[{Aaboud et~al.(2018)}]{ATLAS:2018hqe}
\bibinfo{author}{M.~Aaboud}, et~al. (\bibinfo{collaboration}{ATLAS}),
\newblock \bibinfo{title}{{Prompt and non-prompt $J/\psi $ and $\psi (2\mathrm
  {S})$ suppression at high transverse momentum in $5.02~\mathrm {TeV}$ Pb+Pb
  collisions with the ATLAS experiment}},
\newblock \bibinfo{journal}{Eur. Phys. J. C} \bibinfo{volume}{78}
  (\bibinfo{year}{2018}) \bibinfo{pages}{762}.
\bibitem[{ALI(2022)}]{ALICE:2022jeh}
\bibinfo{title}{{$\psi(2S)$ suppression in Pb-Pb collisions at the LHC}},
\newblock \bibinfo{journal}{arXiv} \bibinfo{volume}{2210.08893}
  (\bibinfo{year}{2022}).
\bibitem[{Zhao and Rapp(2010)}]{Zhao:2010nk}
\bibinfo{author}{X.~Zhao}, \bibinfo{author}{R.~Rapp},
\newblock \bibinfo{title}{{Charmonium in Medium: From Correlators to
  Experiment}},
\newblock \bibinfo{journal}{Phys. Rev. C} \bibinfo{volume}{82}
  (\bibinfo{year}{2010}) \bibinfo{pages}{064905}.
\bibitem[{Du et~al.(2017)Du, Rapp, and He}]{Du:2017qkv}
\bibinfo{author}{X.~Du}, \bibinfo{author}{R.~Rapp}, \bibinfo{author}{M.~He},
\newblock \bibinfo{title}{{Color screening and regeneration of bottomonia in
  high-energy heavy-ion collisions}},
\newblock \bibinfo{journal}{Phys. Rev. C} \bibinfo{volume}{96}
  (\bibinfo{year}{2017}) \bibinfo{pages}{054901}.
\bibitem[{Ferreiro et~al.(2010)Ferreiro, Fleuret, Lansberg, and
  Rakotozafindrabe}]{Ferreiro:2009ur}
\bibinfo{author}{E.~G. Ferreiro}, \bibinfo{author}{F.~Fleuret},
  \bibinfo{author}{J.~P. Lansberg}, \bibinfo{author}{A.~Rakotozafindrabe},
\newblock \bibinfo{title}{{Centrality, rapidity and transverse-Momentum
  dependence of cold nuclear matter effects on $J/\psi$ production in dAu, CuCu
  and AuAu collisions at $\sqrt{s_{_\mathrm{NN}}}$ = 200 GeV}},
\newblock \bibinfo{journal}{Phys. Rev. C} \bibinfo{volume}{81}
  (\bibinfo{year}{2010}) \bibinfo{pages}{064911}.
\bibitem[{Arleo and Peign\'e(2014)}]{Arleo:2014oha}
\bibinfo{author}{F.~Arleo}, \bibinfo{author}{S.~Peign\'e},
\newblock \bibinfo{title}{{Quarkonium suppression in heavy-ion collisions from
  coherent energy loss in cold nuclear matter}},
\newblock \bibinfo{journal}{JHEP} \bibinfo{volume}{10} (\bibinfo{year}{2014})
  \bibinfo{pages}{073}.
\bibitem[{Gavin and Vogt(1997)}]{Gavin:1996yd}
\bibinfo{author}{S.~Gavin}, \bibinfo{author}{R.~Vogt},
\newblock \bibinfo{title}{{Charmonium suppression by Comover scattering in
  Pb+Pb collisions}},
\newblock \bibinfo{journal}{Phys. Rev. Lett.} \bibinfo{volume}{78}
  (\bibinfo{year}{1997}) \bibinfo{pages}{1006--1009}.
\bibitem[{Adamczyk et~al.(2014)}]{STAR:2013kwk}
\bibinfo{author}{L.~Adamczyk}, et~al. (\bibinfo{collaboration}{STAR}),
\newblock \bibinfo{title}{{Suppression of $\Upsilon$ production in d+Au and
  Au+Au collisions at $\sqrt{s_{_\mathrm{NN}}}$ = 200 GeV}},
\newblock \bibinfo{journal}{Phys. Lett. B} \bibinfo{volume}{735}
  (\bibinfo{year}{2014}) \bibinfo{pages}{127--137}. \bibinfo{note}{[Erratum:
  Phys.Lett.B 743, 537--541 (2015)]}.
\bibitem[{Khachatryan et~al.(2017)}]{CMS:2016rpc}
\bibinfo{author}{V.~Khachatryan}, et~al. (\bibinfo{collaboration}{CMS}),
\newblock \bibinfo{title}{{Suppression of $\Upsilon(1S), \Upsilon(2S)$ and
  $\Upsilon(3S)$ production in PbPb collisions at $\sqrt{s_{_\mathrm{NN}}}$ =
  2.76 TeV}},
\newblock \bibinfo{journal}{Phys. Lett. B} \bibinfo{volume}{770}
  (\bibinfo{year}{2017}) \bibinfo{pages}{357--379}.
\bibitem[{Sirunyan et~al.(2019)}]{CMS:2018zza}
\bibinfo{author}{A.~M. Sirunyan}, et~al. (\bibinfo{collaboration}{CMS}),
\newblock \bibinfo{title}{{Measurement of nuclear modification factors of
  $\Upsilon$(1S), $\Upsilon$(2S), and $\Upsilon$(3S) mesons in PbPb collisions
  at $\sqrt{s_{_\mathrm{NN}}} =$ 5.02 TeV}},
\newblock \bibinfo{journal}{Phys. Lett. B} \bibinfo{volume}{790}
  (\bibinfo{year}{2019}) \bibinfo{pages}{270--293}.
\bibitem[{Acharya et~al.(2021)}]{ALICE:2020wwx}
\bibinfo{author}{S.~Acharya}, et~al. (\bibinfo{collaboration}{ALICE}),
\newblock \bibinfo{title}{{$\Upsilon$ production and nuclear modification at
  forward rapidity in Pb-Pb collisions at $\sqrt{s_{_\mathrm{NN}}}$ = 5.02
  TeV}},
\newblock \bibinfo{journal}{Phys. Lett. B} \bibinfo{volume}{822}
  (\bibinfo{year}{2021}) \bibinfo{pages}{136579}.
\bibitem[{Adamczyk et~al.(2016)}]{Adamczyk:2016dzv}
\bibinfo{author}{L.~Adamczyk}, et~al. (\bibinfo{collaboration}{STAR}),
\newblock \bibinfo{title}{{$\Upsilon$ production in U+U collisions at
  $\sqrt{s_{_\mathrm{NN}}}$ = 193 GeV measured with the STAR experiment}},
\newblock \bibinfo{journal}{Phys. Rev. C} \bibinfo{volume}{94}
  (\bibinfo{year}{2016}) \bibinfo{pages}{064904}.
\bibitem[{Ackermann et~al.(2003)}]{Ackermann:2002ad}
\bibinfo{author}{K.~Ackermann}, et~al. (\bibinfo{collaboration}{STAR}),
\newblock \bibinfo{title}{{STAR detector overview}},
\newblock \bibinfo{journal}{Nucl. Instrum. Meth. A} \bibinfo{volume}{499}
  (\bibinfo{year}{2003}) \bibinfo{pages}{624--632}.
\bibitem[{Anderson et~al.(2003)}]{Anderson:2003ur}
\bibinfo{author}{M.~Anderson}, et~al.,
\newblock \bibinfo{title}{{The Star time projection chamber: A Unique tool for
  studying high multiplicity events at RHIC}},
\newblock \bibinfo{journal}{Nucl. Instrum. Meth. A} \bibinfo{volume}{499}
  (\bibinfo{year}{2003}) \bibinfo{pages}{659--678}.
\bibitem[{Beddo et~al.(2003)}]{Beddo:2002zx}
\bibinfo{author}{M.~Beddo}, et~al. (\bibinfo{collaboration}{STAR}),
\newblock \bibinfo{title}{{The STAR barrel electromagnetic calorimeter}},
\newblock \bibinfo{journal}{Nucl. Instrum. Meth. A} \bibinfo{volume}{499}
  (\bibinfo{year}{2003}) \bibinfo{pages}{725--739}.
\bibitem[{Ruan et~al.(2009)}]{Ruan:2009ug}
\bibinfo{author}{L.~Ruan}, et~al.,
\newblock \bibinfo{title}{{Perspectives of a Midrapidity Dimuon Program at
  RHIC: A Novel and Compact Muon Telescope Detector}},
\newblock \bibinfo{journal}{J. Phys. G} \bibinfo{volume}{36}
  (\bibinfo{year}{2009}) \bibinfo{pages}{095001}.
\bibitem[{Adam et~al.(2019)}]{STAR:2019fge}
\bibinfo{author}{J.~Adam}, et~al. (\bibinfo{collaboration}{STAR}),
\newblock \bibinfo{title}{{Measurement of inclusive $J/\psi$ suppression in
  Au+Au collisions at $\sqrt{s_{_\mathrm{NN}}}$ = 200 GeV through the dimuon
  channel at STAR}},
\newblock \bibinfo{journal}{Phys. Lett. B} \bibinfo{volume}{797}
  (\bibinfo{year}{2019}) \bibinfo{pages}{134917}.
\bibitem[{Huang et~al.(2016)}]{Huang:2016dbm}
\bibinfo{author}{T.~C. Huang}, et~al.,
\newblock \bibinfo{title}{{Muon Identification with Muon Telescope Detector at
  the STAR Experiment}},
\newblock \bibinfo{journal}{Nucl. Instrum. Meth. A} \bibinfo{volume}{833}
  (\bibinfo{year}{2016}) \bibinfo{pages}{88--93}.
\bibitem[{Miller et~al.(2007)Miller, Reygers, Sanders, and
  Steinberg}]{Miller:2007ri}
\bibinfo{author}{M.~L. Miller}, \bibinfo{author}{K.~Reygers},
  \bibinfo{author}{S.~J. Sanders}, \bibinfo{author}{P.~Steinberg},
\newblock \bibinfo{title}{{Glauber modeling in high energy nuclear
  collisions}},
\newblock \bibinfo{journal}{Ann. Rev. Nucl. Part. Sci.} \bibinfo{volume}{57}
  (\bibinfo{year}{2007}) \bibinfo{pages}{205--243}.
\bibitem[{Brun et~al.(1987)Brun, Bruyant, Maire, McPherson, and
  Zanarini}]{Brun:1987ma}
\bibinfo{author}{R.~Brun}, \bibinfo{author}{F.~Bruyant},
  \bibinfo{author}{M.~Maire}, \bibinfo{author}{A.~C. McPherson},
  \bibinfo{author}{P.~Zanarini},
\newblock \bibinfo{title}{{GEANT3}},
\newblock \bibinfo{journal}{CERN-DD-EE-84-1}  (\bibinfo{year}{1987}).
\bibitem[{Sjostrand et~al.(2006)Sjostrand, Mrenna, and
  Skands}]{Sjostrand:2006za}
\bibinfo{author}{T.~Sjostrand}, \bibinfo{author}{S.~Mrenna},
  \bibinfo{author}{P.~Z. Skands},
\newblock \bibinfo{title}{{PYTHIA 6.4 Physics and Manual}},
\newblock \bibinfo{journal}{JHEP} \bibinfo{volume}{05} (\bibinfo{year}{2006})
  \bibinfo{pages}{026}.
\bibitem[{Feldman and Cousins(1998)}]{Feldman:1997qc}
\bibinfo{author}{G.~J. Feldman}, \bibinfo{author}{R.~D. Cousins},
\newblock \bibinfo{title}{{Unified approach to the classical statistical
  analysis of small signals}},
\newblock \bibinfo{journal}{Phys. Rev. D} \bibinfo{volume}{57}
  (\bibinfo{year}{1998}) \bibinfo{pages}{3873--3889}.
\bibitem[{Abelev et~al.(2010)}]{Abelev:2010am}
\bibinfo{author}{B.~Abelev}, et~al. (\bibinfo{collaboration}{STAR}),
\newblock \bibinfo{title}{{$\Upsilon$ cross section in $p+p$ collisions at
  $\sqrt{s} = 200$ GeV}},
\newblock \bibinfo{journal}{Phys. Rev. D} \bibinfo{volume}{82}
  (\bibinfo{year}{2010}) \bibinfo{pages}{012004}.
\bibitem[{Adare et~al.(2015)}]{PHENIX:2014tbe}
\bibinfo{author}{A.~Adare}, et~al. (\bibinfo{collaboration}{PHENIX}),
\newblock \bibinfo{title}{{Measurement of $\Upsilon(1S+2S+3S)$ production in
  $p+p$ and Au$+$Au collisions at $\sqrt{s_{_\mathrm{NN}}}$ = 200 GeV}},
\newblock \bibinfo{journal}{Phys. Rev. C} \bibinfo{volume}{91}
  (\bibinfo{year}{2015}) \bibinfo{pages}{024913}.
\bibitem[{Zha et~al.(2013)Zha, Yang, Huang, Ruan, Yang, Tang, and
  Xu}]{Zha:2013uoa}
\bibinfo{author}{W.~Zha}, \bibinfo{author}{C.~Yang},
  \bibinfo{author}{B.~Huang}, \bibinfo{author}{L.~Ruan},
  \bibinfo{author}{S.~Yang}, \bibinfo{author}{Z.~Tang},
  \bibinfo{author}{Z.~Xu},
\newblock \bibinfo{title}{{Systematic study of the experimental measurements on
  ratios of different $\Upsilon$ states}},
\newblock \bibinfo{journal}{Phys. Rev. C} \bibinfo{volume}{88}
  (\bibinfo{year}{2013}) \bibinfo{pages}{067901}.
\bibitem[{Zhu et~al.(2008)}]{NuSea:2007ult}
\bibinfo{author}{L.~Y. Zhu}, et~al. (\bibinfo{collaboration}{NuSea}),
\newblock \bibinfo{title}{{Measurement of $\Upsilon$ Production for $p+p$ and
  $p+d$ Interactions at 800 GeV/c}},
\newblock \bibinfo{journal}{Phys. Rev. Lett.} \bibinfo{volume}{100}
  (\bibinfo{year}{2008}) \bibinfo{pages}{062301}.
\bibitem[{Acosta et~al.(2002)}]{CDF:2001fdy}
\bibinfo{author}{D.~Acosta}, et~al. (\bibinfo{collaboration}{CDF}),
\newblock \bibinfo{title}{{$\Upsilon$ Production and Polarization in $p\bar{p}$
  Collisions at $\sqrt{s}=$ 1.8 TeV}},
\newblock \bibinfo{journal}{Phys. Rev. Lett.} \bibinfo{volume}{88}
  (\bibinfo{year}{2002}) \bibinfo{pages}{161802}.
\bibitem[{Chatrchyan et~al.(2013)}]{CMS:2013qur}
\bibinfo{author}{S.~Chatrchyan}, et~al. (\bibinfo{collaboration}{CMS}),
\newblock \bibinfo{title}{{Measurement of the $\Upsilon(1S), \Upsilon(2S)$, and
  $\Upsilon(3S)$ cross sections in $pp$ collisions at $\sqrt{s}$ = 7 TeV}},
\newblock \bibinfo{journal}{Phys. Lett. B} \bibinfo{volume}{727}
  (\bibinfo{year}{2013}) \bibinfo{pages}{101--125}.
\bibitem[{Lansberg(2020)}]{Lansberg:2019adr}
\bibinfo{author}{J.-P. Lansberg},
\newblock \bibinfo{title}{{New Observables in Inclusive Production of
  Quarkonia}},
\newblock \bibinfo{journal}{Phys. Rept.} \bibinfo{volume}{889}
  (\bibinfo{year}{2020}) \bibinfo{pages}{1--106}.
\bibitem[{Aaboud et~al.(2018)}]{ATLAS:2017prf}
\bibinfo{author}{M.~Aaboud}, et~al. (\bibinfo{collaboration}{ATLAS}),
\newblock \bibinfo{title}{{Measurement of quarkonium production in
  proton\textendash{}lead and proton\textendash{}proton collisions at
  $5.02~\mathrm {TeV}$ with the ATLAS detector}},
\newblock \bibinfo{journal}{Eur. Phys. J. C} \bibinfo{volume}{78}
  (\bibinfo{year}{2018}) \bibinfo{pages}{171}.
\bibitem[{Tumasyan et~al.(2022)}]{CMS:2022wfi}
\bibinfo{author}{A.~Tumasyan}, et~al. (\bibinfo{collaboration}{CMS}),
\newblock \bibinfo{title}{{Nuclear modification of $\Upsilon$ states in pPb
  collisions at $\sqrt{s_\mathrm{NN}}$ = 5.02 TeV}},
\newblock \bibinfo{journal}{arXiv} \bibinfo{volume}{2202.11807}
  (\bibinfo{year}{2022}).
\bibitem[{Brambilla et~al.(2021{\natexlab{a}})Brambilla, Escobedo, Strickland,
  Vairo, Vander~Griend, and Weber}]{Brambilla:2021wkt}
\bibinfo{author}{N.~Brambilla}, \bibinfo{author}{M.~A. Escobedo},
  \bibinfo{author}{M.~Strickland}, \bibinfo{author}{A.~Vairo},
  \bibinfo{author}{P.~Vander~Griend}, \bibinfo{author}{J.~H. Weber},
\newblock \bibinfo{title}{{Bottomonium production in heavy-ion collisions using
  quantum trajectories: Differential observables and momentum anisotropy}},
\newblock \bibinfo{journal}{Phys. Rev. D} \bibinfo{volume}{104}
  (\bibinfo{year}{2021}{\natexlab{a}}) \bibinfo{pages}{094049}.
\bibitem[{Brambilla et~al.(2021{\natexlab{b}})Brambilla, Escobedo, Strickland,
  Vairo, Vander~Griend, and Weber}]{Brambilla:2020qwo}
\bibinfo{author}{N.~Brambilla}, \bibinfo{author}{M.~A. Escobedo},
  \bibinfo{author}{M.~Strickland}, \bibinfo{author}{A.~Vairo},
  \bibinfo{author}{P.~Vander~Griend}, \bibinfo{author}{J.~H. Weber},
\newblock \bibinfo{title}{{Bottomonium suppression in an open quantum system
  using the quantum trajectories method}},
\newblock \bibinfo{journal}{JHEP} \bibinfo{volume}{05}
  (\bibinfo{year}{2021}{\natexlab{b}}) \bibinfo{pages}{136}.
\bibitem[{Brambilla et~al.(2022)Brambilla, Escobedo, Islam, Strickland, Tiwari,
  Vairo, and Vander~Griend}]{Brambilla:2022ynh}
\bibinfo{author}{N.~Brambilla}, \bibinfo{author}{M.~A. Escobedo},
  \bibinfo{author}{A.~Islam}, \bibinfo{author}{M.~Strickland},
  \bibinfo{author}{A.~Tiwari}, \bibinfo{author}{A.~Vairo},
  \bibinfo{author}{P.~Vander~Griend},
\newblock \bibinfo{title}{{Heavy quarkonium dynamics at next-to-leading order
  in the binding energy over temperature}},
\newblock \bibinfo{journal}{arXiv} \bibinfo{volume}{2205.10289}
  (\bibinfo{year}{2022}).
\bibitem[{Yao et~al.(2021)Yao, Ke, Xu, Bass, and M\"uller}]{Yao:2020xzw}
\bibinfo{author}{X.~Yao}, \bibinfo{author}{W.~Ke}, \bibinfo{author}{Y.~Xu},
  \bibinfo{author}{S.~A. Bass}, \bibinfo{author}{B.~M\"uller},
\newblock \bibinfo{title}{{Coupled Boltzmann Transport Equations of Heavy
  Quarks and Quarkonia in Quark-Gluon Plasma}},
\newblock \bibinfo{journal}{JHEP} \bibinfo{volume}{01} (\bibinfo{year}{2021})
  \bibinfo{pages}{046}.
\bibitem[{Hoelck et~al.(2017)Hoelck, Nendzig, and Wolschin}]{Hoelck:2016tqf}
\bibinfo{author}{J.~Hoelck}, \bibinfo{author}{F.~Nendzig},
  \bibinfo{author}{G.~Wolschin},
\newblock \bibinfo{title}{{In-medium $\Upsilon$ suppression and feed-down in UU
  and PbPb collisions}},
\newblock \bibinfo{journal}{Phys. Rev. C} \bibinfo{volume}{95}
  (\bibinfo{year}{2017}) \bibinfo{pages}{024905}.

\end{thebibliography}

\end{document}